\documentclass[twocolumn,preprintnumbers,amsmath,amssymb,showpacs,prb,aps]{revtex4}
\usepackage{graphicx}
\usepackage{dcolumn}
\usepackage{amsmath}
\usepackage[perpage,symbol*]{footmisc}

\makeatletter
\def\btt#1{\texttt{\@backslashchar#1}}%
\DeclareRobustCommand\bblash{\btt{\@backslashchar}}%
\makeatother

\begin{document}
\draft \widetext

\title{Dynamics of a one-dimensional Holstein polaron with Davydov Ans\"{a}tze }

\author{Jin Sun, Bin Luo, and Yang Zhao\footnote{Electronic address:~\url{YZhao@ntu.edu.sg}}}

\affiliation{\it School of Materials Science and Engineering,
Nanyang Technological University, Singapore 639798}
\date{\today}
\widetext

\begin{abstract}
Following the Dirac-Frenkel time-dependent variational principle, dynamics of a one-dimensional Holstein polaron is probed by employing the Davydov ${\rm D}_2$ Ansatz with two sets of variational parameters, one for each constituting particle in the exciton-phonon system, and a simplified variant of the Davydov ${\rm D}_1$ Ansatz, also known as the $\tilde{\rm D}$ Ansatz, with an additional set of phonon displacement parameters. A close examination of variational outputs from the two trial states reveals fine details  of the polaron structure and intricacies of dynamic exciton-phonon interactions. Superradiance coherence sizes, speeds of exciton-induced phonon wave packets, linear optical absorption, and polaron energy compositions are also included in the study.
\end{abstract}

\maketitle \narrowtext

\section{Introduction}
An electron in an insulating crystal induces a local lattice
distortion around itself as it is excited by a photon. The electron
together with the locally distorted lattice around it can be
viewed as a quasiparticle which is also known as a polaron.
Relaxation dynamics of photoexcited entities such as polarons in
liquids and solids has recently received much interest thanks to the
advent of the ultrafast laser spectroscopy
\cite{Dynamic3,Dynamic4,Dynamic5}. It is now
commonly accepted that dephasing and relaxation time scales in
condensed matter are approximately picoseconds to tens of
picoseconds. Emerging technological capabilities to control
femtosecond pulse durations and down-to-one-hertz bandwidth
resolutions provide novel probes on vibrational dynamics and
excitation relaxation. For example, progress in femtosecond
spectroscopic techniques has made it possible to observe a coherent
phonon wave packet oscillating along an adiabatic potential surface
associated with a self-trapped exciton in a crystal with strong
exciton-phonon interactions \cite{Tomimoto}.

Ultrafast events occur on femtosecond to picosecond
timescales, and studies of ultrafast relaxation dynamics are a
strategic research domain from both fundamental and technological
points of view. It is the aim of an ultrafast optical experiment to
provide information on the details of temporal evolution on a
femtosecond scale, which, in turn, offers insights into fundamental
processes governing the dynamics. Developments in ultrafast laser
physics and technologies now allow studies of nonequilibrium
carrier/exciton dynamics that is previously inaccessible to
traditional linear optical spectroscopy. However, theoretical studies of
polaron dynamics has not received much-deserved attention due to inherent
difficulties in obtaining reliable solutions.
Previously, a time-dependent form of the Merrifield-type polaron wave function
with zero crystal momentum has been employed to yield an approximative solution to
the Schr\"{o}dinger equation that governs the ultrafast relaxation
process of a one-dimensional molecular chain
\cite{merr,merr_zhao,Dynamic1,Dynamic2}. Results show that temporal
changes of the exciton coherence size and related energy relaxation
strongly depend on the exciton transfer integral, the exciton-phonon
coupling strength, and the phonon bandwidth. The applicability
of the Merrifield wave function, however, is restricted to the narrow-band
regime where the electronic coupling between neighboring molecules is
sufficiently weak leaving the electron-phonon coupling at a dominant
role. In addition, in the presence of off-diagonal electron-phonon
interactions, the Merrifield Ansatz is shown to fail
\cite{toyozawa2}.

Beyond the Merrifield Ansatz, there exist several trial wave
functions of increasing sophistication to describe the polaron state
in a translationally invariant manner such as the Toyozawa Ansatz \cite{toyozawa0},
the Global-Local (GL) Ansatz formulated by Zhao and coworkers in the
early 90s \cite{thesis,toyozawa1}, and a delocalized form of the
Davydov ${\rm D}_1$ Ansatz that has been constructed very recently
\cite{zhaoetal}. By using these Ans\"{a}tze, we have previously
investigated the ground state polaron energy band and the self-trapping phenomenon
of a static Holstein polaron. Far superior results
have been obtained compared with the Merrifield Ansatz
\cite{toyozawa1,toyozawa2,GL2}. Closely related to those polaron
trial states are the Davydov Ans\"{a}tze which originated from the theory
of ``Davydov soliton." Seeking to explain storage and transport of
biological energy in protein structures, Davydov proposed in 1973
that quantum units of peptide vibrational energy might become
``self-localized" through interactions with lattice phonons \cite{Davydov0, Davydov0_2}.
Following his suggestion, many related studies have been
carried out on the ``Davydov soliton," an essentially one-dimensional object
that maintains dynamic integrity by balancing the effects
of nonlinearity against those of dispersion.
The original Davydov Ans\"{a}tze include two forms of varying sophistication, namely,
the ${\rm D}_1$
\cite{Davydov1, Davydov1_2, Davydov1_3, Davydov1_4, Davydov1_5, Davydov1_6, Davydov1_7}
and ${\rm D}_2$ Ans\"{a}tze,
with the latter being a special case of the former.

In this work, we employ the Davydov ${\rm D}_2$ Ansatz and a localized form factor
of the GL Ansatz, previously known as the $\tilde{\rm D}$ Ansatz, to study
time evolution of the Holstein polaron following
the Dirac-Frenkel time-dependent variational approach\cite{Dirac}.
Time-dependent variational parameters which specify the two
Ans\"{a}tze are obtained from solving a set of coupled differential
equations generated by the Lagrangian formalism of the Dirac-Frenkel variation.
Special attention is paid to the evolution of
the reduced single-exciton density matrix and the propagation of exciton and phonon amplitudes
from an initial location, where the exciton resides at $t=0$, to the entire aggregate.

The paper is organized as follows. In Sec.~II we introduce the model
Hamiltonian and discuss the nature of the trial
states we use for dynamics studies, which is followed by the procedure
of the time-dependent variation. In
Sec.~III results from our dynamics calculation using the two trial states
are displayed and discussed. Conclusions are drawn in Sec. IV.

\section{Methodology}

We consider a one-dimensional aggregate of N molecules with
a periodic boundary condition. There is only one two-level electronic
system for each molecule coupled linearly with the phonon field.
The Holstein Hamiltonian for the exciton-phonon system can be written as
\cite{Frolich,Holstein,Mahan}
\begin{equation}
\hat{H}=\hat{H}_{\rm ex}+\hat{H}_{\rm ex-ph}+\hat{H}_{\rm ph} \label{Hamiltonian}
\end{equation}
with
\begin{equation}
\hat{H}_{\rm ex}=-J\sum_{n}\hat{B}_{n}^{\dagger}(\hat{B}_{n+1}+\hat{B}_{n-1}) \label{H_ex}
\end{equation}
\begin{equation}
\hat{H}_{\rm ph}=\sum_{q}\omega_{q}\hat{b}_{q}^{\dagger}\hat{b}_{q} \label{H_ph}
\end{equation}
\begin{equation}
\hat{H}_{\rm ex-ph}=\sum_{q,n=1}^{N}g_{q}\omega_{q}\hat{B}_{n}^{\dagger}\hat{B}_{n}(\hat{b}_{q}
e^{iqn}+\hat{b}_{q}^{\dagger}e^{-iqn})
\label{ex-ph}
\end{equation}
Here $\hat{H}_{\rm ex}$ is the Hamiltonian for a single Frenkel exciton
band in a rigid chain, and $\hat{B}_{n} (\hat{B}_{n}^{\dagger})$ is the
Pauli annihilation (creation) operator of an exciton at the $n$th
site. We set $\hbar=1$ and assume the nearest-neighbor exciton
transfer integral $J_{mn}=J\delta_{m,n\pm1}$. $\hat{H}_{\rm ph}$ is the phonon
Hamiltonian where $\hat{b}_{q} (\hat{b}_{q}^{\dagger})$ is the boson
annihilation (creation) operator of a phonon with monentum $q$ and
frequency $\omega_{q}$. For simplicity, the zero-point energy is
neglected. $\hat{H}_{\rm ex-ph}$ assumes the exciton is coupled
linearly with the phonon field in a site diagonal form.

Because there is no exciton in the ground state, the Hamiltonian for
the ground state is represented by
\begin{equation}
\hat{H}_{\rm g}=|0\rangle_{\rm ex}~\hat{H}_{\rm ph}\ _{\rm ex}\langle0|
\end{equation}
where $|0\rangle_{\rm ex}$ stands for the exciton vacuum. The global
ground state is then described as a direct product of both vacuum
states of the exciton and the phonon field: $|G\rangle=|0\rangle_{\rm ex}|0\rangle_{\rm ph}$.
We confine ourselves to one-exciton subspace for the optically
excited state \cite{Lu} since the exciton number is conserved in the
total Hamiltonian in Eq.~(\ref {Hamiltonian}).

The spectral density \cite{Mukamel} embodying all relevant information of
the coupled exciton-phonon system in Eq.~(\ref {ex-ph}) can be written as
\begin{equation}
C_{mn}(\omega)\equiv\frac{1}{2\pi}\int_{-\infty}^{\infty}
\langle\hat{V}_{m}(t)\hat{V}_{n}(0)\rangle_{\rm
ph}e^{i\omega t}dt \label{spectral-density}
\end{equation}
where $\hat{V}_{n}(t)$ is a Heisenberg representation of the
exciton-phonon interaction at the $n$th site
\begin{equation}
\hat{V}_{n}=\sum_{q}g_{q}\omega_{q}e^{iqn}(\hat{b}_{q}+\hat{b}_{-q}^{\dagger})
\label{interaction}
\end{equation}
and the $\langle\cdots\rangle_{\rm ph}$ denotes the thermal average in
the free-phonon basis. Note that $C_{mn}(\omega)$
represents not only the time correlation but also the spatial
correlation for the exction-phonon interactions \cite{Mukamel}. Here
only the case of zero temperature is considered. Substituting
Eq.~(\ref{interaction}) into Eq.~(\ref{spectral-density}), one has
\begin{equation}
C_{mn}(\omega)=\sum_{q}g_{q}^{2}\omega_{q}^{2}e^{iq(m-n)}\delta(\omega-\omega_{q})
\end{equation}
Following Tanaka \cite{Dynamic2}, the spectral density is assumed to have the form
\begin{eqnarray}
C_{00}(\omega)&=&\sum_{q}g_{q}^{2}\omega_{q}^{2}\delta(\omega-\omega_{q}) \nonumber \\
&=&\frac{2 S \omega^2 }{\pi W^2} \sqrt {W^{2}-(\omega-\omega_{0})^{2}}
\label{C00}
\end{eqnarray}
The relaxation energy is defined by
\begin{equation} \label{relaxationE}
E_r\equiv\int_{-\infty}^{\infty}\frac{C_{00}(\omega)}{\omega}
d\omega=\sum_{q}g_{q}^{2}\omega_{q}\equiv S\omega_{0}
\end{equation}
where $S$ is the Huang-Rhys factor \cite{Huang}, $\omega_{0}$ is the
central energy of the phonon band, and $W$ is the phonon energy band width .
Here we assume a linear dispersion phonon band with a linear dispersion
\begin{equation}
\omega_{q}=\omega_{0} + 2W(\frac{|q|}{\pi}-\frac{1}{2}) \label{width}
\end{equation}
The central frequency of the phonon band $\omega_{0}$ is taken as
the energy unit, i.e., $\omega_{0}=1$. From
Eqs.~(\ref{C00})-(\ref{width}), we can obtain $g_{q}$ for a given $S$ and $W$.

The Merrifield Ansatz, also known as the small-polaron Ansatz \cite{merr,merr_zhao},
is a simple, effective trial state to describe a coupled exciton-phonon system
in which the exciton transfer integral is small compared with other system energy scales.
There is only one set of variational parameters, the phonon displacements $\beta_n^K (t)$,
or alternatively, its Fourier transform $\beta_q^K (t)$,
that characterizes the Merrifield trial state $|\Psi_{\rm M};K\rangle$ of crystal momentum $K$:
\begin{eqnarray}\label{Merr_Ansatz}
|\Psi_{\rm M};K\rangle &=& N^{-1/2}\sum_ne^{iKn}|n\rangle_{\rm ex} \nonumber\\
&&\times\exp[-\sum_{n_b}(\beta_{n_b-n}^K\hat{b}_{n_b}^\dagger - \beta_{n_b-n}^{K*}\hat{b}_{n_b})]|0\rangle_{\rm ph} \nonumber\\
&=& N^{-1/2}\sum_ne^{iKn}|n\rangle_{\rm ex} \nonumber\\
&&\times\exp[-\sum_{q}(\beta_{q}^Ke^{-iqn}\hat{b}_{q}^\dagger - \beta_{q}^{K*}e^{iqn}\hat{b}_{q})]|0\rangle_{\rm ph} \nonumber\\
\end{eqnarray}
In contrast, Davydov built a theory of enveloped solitons based on the two trial states of localized nature
\begin{equation}
|D_{1}(t)\rangle\equiv\sum_{n}\psi_{n}(t)\hat{B}_{n}^{\dagger}|0\rangle_{\rm
ex}\otimes|\lambda_{n}(t)\rangle
\end{equation}
with
\begin{equation}
|\lambda_{n}(t)\rangle\equiv\exp\{\sum_{q}[\lambda_{qn}(t)
\hat{b}_{q}^{\dagger}-\lambda_{qn}^{*}(t)\hat{b}_{q}]\}|0\rangle_{\rm
ph}
\end{equation}
and
\begin{equation}\label{D2def}
|D_{2}(t)\rangle\equiv\sum_{n}\psi_{n}(t)\hat{B}_{n}^{\dagger}|0\rangle_{\rm
ex}\otimes|\lambda(t)\rangle
\end{equation}
with
\begin{equation}\label{D2lambdadef}
|\lambda(t)\rangle\equiv\exp\{\sum_{q}[\lambda_{q}(t)
\hat{b}_{q}^{\dagger}-\lambda_{q}^{*}(t)\hat{b}_{q}]\}|0\rangle_{\rm
ph}
\end{equation}
The ${\rm D}_1$ Ansatz differs from the more widely used ${\rm D}_2$ Ansatz.
In the ${\rm D}_2$ Ansatz the phonon amplitudes are exciton-site independent,
i.e., $\lambda_{qn}(t)=\lambda_{q}(t)$ for all n. The ${\rm D}_2$ Ansatz reprsents
the quantum state of each phonon mode by a single coherent state,
and hence has at all times a strongly classical character. In such solutions,
the typically pulse-shaped distribution of the amplitudes $\lambda_{q}$ describes
a lattice deformation fixed in the frame of the lattice, to which
conforms a correspondingly pulse-shaped distribution of exciton
amplitudes $\psi_{n}$.

Since the description of phonons in the ${\rm D}_2$ Ansatz is unsophisticated,
direct correlations between the amplitudes of phonons and exciton are neglected.
In many cases, the ${\rm D}_2$ Ansatz is inadequate to capture the complexities of the exciton-phonon system.
Therefore, we introduce the $\tilde{\rm D}$ Ansatz, i.e., the localized form factor of the GL Ansatz:
\begin{eqnarray} \label{DtildeDef}
|\tilde{D}(t)\rangle &=&
\sum_{n_{1}}\psi_{n_{1}}(t)\hat{B}_{n}^{\dagger}\exp[\sum_{n_{2}}(\lambda_{n_{2}}-\beta_{n_{2}-n_{1}})\hat{b}_{n_{2}}^{\dag}
\nonumber   \\
&&-(\lambda_{n_{2}}^{*}-\beta_{n_{2}-n_{1}}^{*})\hat{b}_{n_{2}}]|0\rangle     \nonumber   \\
&=&\sum_{n}\psi_{n}(t)\hat{B}_{n}^{\dagger}\exp[N^{-1/2}\sum_{q}(\beta_{q}e^{-iqn}-\lambda_{q})\hat{b}_{q}^{\dagger}
\nonumber   \\
&&-(\beta_{q}^{*}e^{iqn}-\lambda_{q}^{*})\hat{b}_{q}]|0\rangle
\end{eqnarray}
The global amplitude $\lambda_{q}$ can be related to the spatial
average of $\lambda_{qn}$ , and the local amplitude
$\beta_{q}e^{-iqn}$ can be viewed as an Ansatz for the spatial
variation of $\lambda_{qn}$ around this mean value.

The time evolution of the photoexcited state in a one-dimensional
molecular aggregate follows the time-dependent
Schr\"{o}dinger equation. As far as the exciton-phonon coupling
is weak, it can be treated perturbatively leading to a nonlinear exciton equation
with the use of the relaxation superoperator. In the strong coupling
case, however, the perturbative method is no longer valid. There are
several approaches to solve the time-dependent Schr\"{o}dinger
equation. In the Hilbert space, the time-dependent wave function
$|\Phi(t)\rangle$ for the Hamiltonian $\hat{H}$ is parameterized by
$\alpha_{m}(t)$ ($m=1,...,M$):
\begin{equation}
|\Phi(t)\rangle\equiv|\{\alpha_{m}(t)\}\rangle.
\end{equation}
Assume that $|\Phi(t)\rangle$ satisfies the time-dependent
Schr\"{o}dinger equation
\begin{equation}
i\frac{\partial}{\partial t}|\Phi(t)\rangle=\hat{H}|\Phi(t)\rangle.
\label{TD-equation}
\end{equation}
Explicitly putting in the Hamiltonian $\hat{H}$ of Eq.~(\ref{Hamiltonian}) and writing
$\partial|\Phi(t)\rangle/\partial t$ in terms of $\alpha_{m}(t)$ and
their time-derivatives $\dot{\alpha}_{m}(t)$ ($m=1,...,M$), one has
\begin{equation}
i\frac{\partial}{\partial t}|\{\alpha_{m}(t)\}\rangle=\hat{H}|\{\alpha_{m}(t)\}\rangle\label{TD-equation2}
\end{equation}
Projecting Eq.~(\ref{TD-equation2}) onto $M$ different states
$|\Psi_{m}\rangle$ ($m=1,...,M$), one obtains $M$ equations of motion
for the parameters $\alpha_{m}(t)$:
\begin{equation}
\langle\Psi_{m}|\{\alpha_{m}(t)\},\{\dot{\alpha}_{m}(t)\}\rangle=0 \label {projection}
\end{equation}

The approach we adopt in this work is the Lagrangian formalism of the Dirac-Frenkel
time-dependent variational method \cite{Itzykson}, a powerful technique to obtain
approximate dynamics of many-body quantum systems for which exact solutions often elude
researchers. We formulate the Lagrangian $L$ as follows
\begin{equation}
L=\langle\Phi(t)|{\frac{i\hbar}{2}}\frac{\overset{\longleftrightarrow}{\partial}}{\partial
t}-\hat{H}|\Phi(t)\rangle
\end{equation}
From this Lagrangian, equations of motion for $\alpha_{m}(t)$ and their
time-derivatives $\dot{\alpha}_{m}(t)$ ($m=1,...,M$) can then be
obtained
\begin{equation}
\frac{d}{dt}(\frac{\partial
L}{\partial\dot{\alpha_{m}^{*}}})-\frac{\partial
L}{\partial\alpha_{m}^{*}}=0 \label {Lagrangian}
\end{equation}
It can be shown that with a properly chosen set of $|\Psi_{m}\rangle$, Eqs.~(\ref{projection})
and (\ref{Lagrangian}) can coincide. The reader is referred to Appendices \ref{a0},
\ref{a1} and \ref{a2} where details of derivation together with these $M$ equations
for ${\rm D}_2$, $\tilde{\rm D}$ and ${\rm D}_1$ Ans\"{a}tze are given, respectively.
Descriptions of the numerical procedure can be found in Appendix \ref{a3}.

\section{Results and Discussions}

\subsection{Reduced Exciton Density Matrix and Coherence Size}

\begin{figure}[tb]
\begin{center}
\includegraphics[scale=0.45]{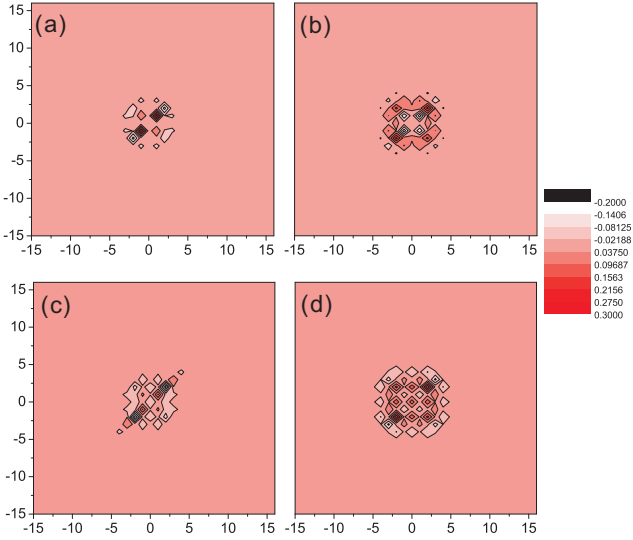}
\end{center}
\caption{The reduced exciton density matrix $\rho_{mn}$ calculated from the $\tilde{\rm D}$ Ansatz for
(a) $t=7\pi/\omega_0$; (b) $t=8\pi/\omega_0$; (c) $t=9\pi/\omega_0$; (d) $t=10\pi/\omega_0$.
The control parameters are $J=0.1,W=0.1$ and $S=0.5$.}\label{den}
\end{figure}

The reduced single-exciton density matrix $\rho_{mn}(t)$ can be obtained after solving
the coupled equations of variational parameters
\begin{equation}
\rho_{mn}(t)={\rm {Tr}}[\rho(t)\hat{B}_{m}^{\dagger}\hat{B}_{n}] \label{rho}
\end{equation}
where $\rho(t)=|\Phi(t)\rangle\langle\Phi(t)|$ is the full density matrix at
zero temperature, and $|\Phi(t)\rangle$ is the total polaron wave
function at time $t$ after the photo excitation takes place.
Substituting the detailed form of the ${\rm D}_2$ Ansatz into the
polaron wave function, one obtains
\begin{eqnarray}
\rho_{mn} = \langle D_{2}|\hat{B}_{m}^{\dagger}\hat{B}_{n}|D_{2}\rangle = \psi_{m}^{*}(t)\psi_{n}(t)
\end{eqnarray}
For the ${\rm D}_2$ Ansatz, the reduced exciton density matrix involves only
the exciton amplitude, and does not contain information of the phonon manifold.
In contrast, the more sophisticated trial state, the $\tilde{\rm D}$ Ansatz,
includes the Debye-Waller factor $S_{n,m}$ in the expression of the reduced exciton density matrix:
\begin{eqnarray}
\rho_{mn} = \langle \tilde{D}|\hat{B}_{m}^{\dagger}\hat{B}_{n}|\tilde{D}\rangle = \lambda_{m}^{*}(t)\lambda_{n}(t)S_{m,n}
\end{eqnarray}
where $S_{n,m}$ can be written as
\begin{eqnarray}
S_{n,m}&=&\exp[-\frac{1}{2}N^{-1}\sum_{q}|\beta_{q}e^{-iqn}-\lambda_{q}|^{2}]   \nonumber   \\
&&\exp[N^{-1}\sum_{q}(\beta_{q}^{*}e^{iqn}-\lambda_{q}^{*})(\beta_{q}e^{-iqm}-\lambda_{q})]   \nonumber \\
&&\exp[-1/2N^{-1}\sum_{q}|\beta_{q}e^{-iqm}-\lambda_{q}|^{2}]
\end{eqnarray}

\begin{figure}[tb]
\begin{center}
\includegraphics[scale=0.5]{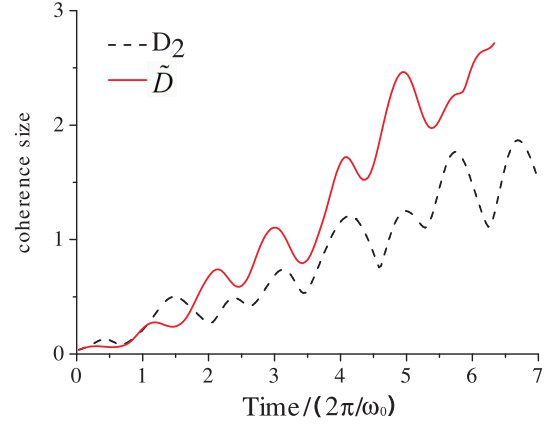}
\end{center}
\caption{The superradiance coherence size of the polaron $L_{\rho}$ for
$J=0.1$, $W=0.1$ and $S=0.5$ calculated from the $\tilde{\rm D}$ (solid) and ${\rm D}_2$ (dashed) Ans\"{a}tze.}\label{coherence}
\end{figure}

Using the Merrifield Ansatz \cite{Dynamic2}, when $J=0$, solutions to
coupled equations of variational parameters can be obtained analytically for
zero phonon bandwidth ($W=0$). It is found that
$L_{\rho}=1/N$ for all $t=2\pi n /\omega_{0}$ ($n=1,2,3,...$).
Periodic changes of the polaron structure as a function of time have also been revealed
when $J$ and $W$ are both small.
At $\omega_0 t =7\pi, 8\pi, 9\pi$ and $10\pi$, the reduced exciton density matrix $\rho_{mn}$
calculated from the $\tilde{\rm D}$ Ansatz
for the case of
$J=0.1, W=0.1$ and $S=0.5$ is displayed in Fig.~\ref{den}.
At $t=0$, $\rho_{mn}$ is confined to the $(0,0)$ point, i.e.,
the exciton is localized at one site of the one-dimensional system.
Then, as the excited state relaxes, the non-zero elements of $\rho_{mn}$ spread
out gradually.
Oscillatory behavior is visible during the evolution of the
density matrix as both the $J$ and $W$ are small, and as a result, dephasing due to phonon dispersion
and exciton transfer is at a minimal level.
When $t$ equals to $2n\pi/\omega_0$, the matrix is
rather delocalized and have a symmetric distribution, and at $ t =(2n+1)\pi/\omega_0$,
matrix elements aggregate along the diagonal line.

When $J$ is increased, the oscillatory behavior of the density matrix gradually fades
as the exciton density delocalization happens more swiftly.
From the reduced exciton density matrix, one can obtain an exciton
coherence size $L_{\rho}$ \cite{Meier,Zhao,Zhao2D} characterizing the
size of a domain within which the chromophores emit coherently.
All information relevant to the excitonic superradiance
is contained within the reduced exciton density matrix $\rho_{mn}(t)$,
the time dependence of which provides a clear view of the relaxation process.
Previously, for translationally invariant states, a
definition of the characteristic coherence size $L_{\rho}$ in terms
of the density matrix was introduced by applying the inverse participation ratio
concept commonly used in the theory of quantum localization:
\begin{equation}
L_{\rho}\equiv[L_{0}\sum_{mn}|\rho_{mn}|^{2}]^{-1}[(\sum_{mn}|\rho_{mn}|)^{2}]
\end{equation}
This definition of the coherence size is borrowed to quantify the excitonic delocalization despite that
our trial states are both of the localized form. Values of $L_{\rho}$ calculated from the $\rm D_2$
and $\tilde{\rm D}$
trial states for the case of
$J=0.1, W=0.1$ and $S=0.5$ are shown in Fig.~\ref{coherence}
as a function of time. At $t=0$, the coherence size
is 1/32 because the exciton is localized to one site completely. Then
coherence size will increase in an oscillatory fashion. When $t$
equals $2n\pi/ \omega_0$, the coherence size will reach a local maximum
and then decrease until $t$ equals $(2n+1)\pi/\omega_0$. This is
consistent with the time evolution of the reduced exciton density matrix.

\begin{figure}[b]
\begin{center}
\includegraphics[scale=0.65]{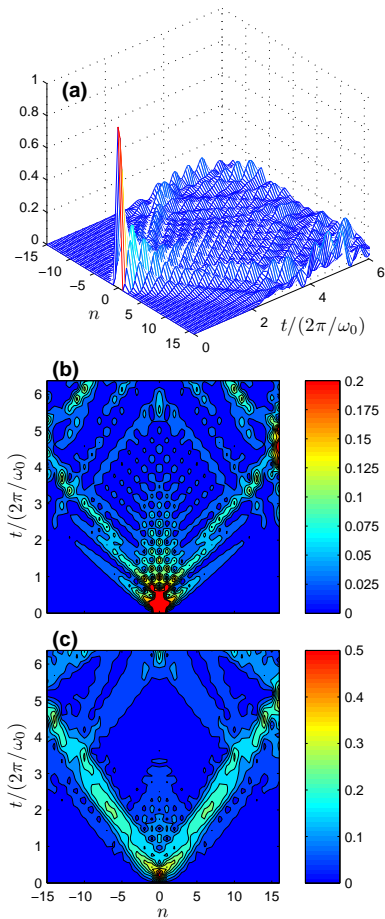}
\end{center}
\caption{Time evolution of variation parameters of the ${\rm D}_2$ Ansatz for $J=0.5,~W=0.8$ and $S=0.5$.
(a) A bird's-eye view of the exciton probability $|\psi_{n}(t)|^{2}$ in real space;
(b) a contour plot of $|\psi_{n}(t)|^{2}$;
(c) a contour plot of the phonon displacement $|\lambda_{n}(t)|$ in real space.} \label{D2-1}
\end{figure}

\begin{figure}[b]
\begin{center}
\includegraphics[scale=0.65]{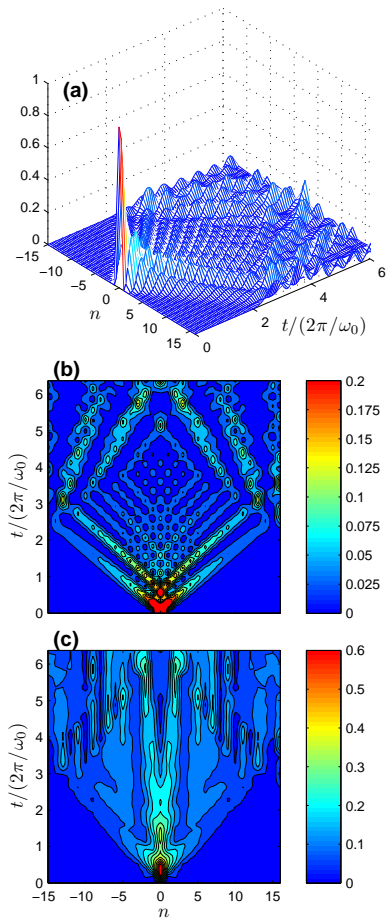}
\end{center}
\caption{Time evolution of variation parameters of the ${\rm D}_2$ Ansatz for $J=0.5,~W=0.1$ and $S=0.5$.
(a) A bird's-eye view of the exciton probability $|\psi_{n}(t)|^{2}$ in real space;
(b) a contour plot of $|\psi_{n}(t)|^{2}$;
(c) a contour plot of the phonon displacement $|\lambda_{n}(t)|$ in real space.} \label{D2-2}
\end{figure}

\subsection{Exciton Amplitudes and Phonon Displacements of the ${\rm D}_2$ Ansatz}

\begin{figure}[tb]
\begin{center}
\includegraphics[scale=0.65]{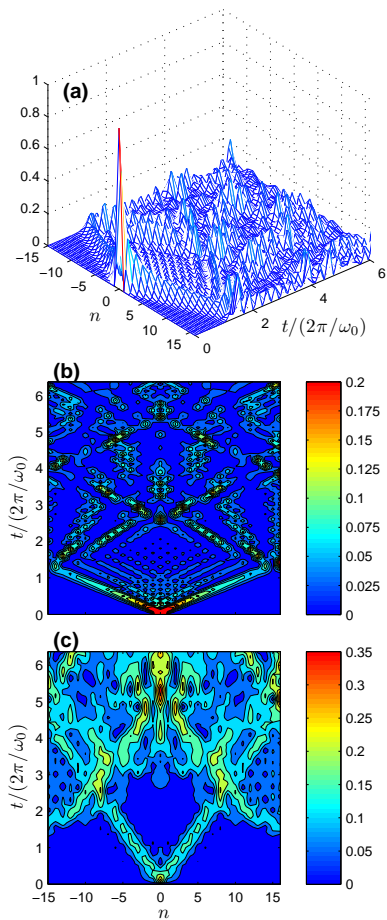}
\end{center}
\caption{Time evolution of variation parameters of the ${\rm D}_2$ Ansatz for $J=1.0,~W=0.8$ and $S=0.5$.
(a) A bird's-eye view of the exciton probability $|\psi_{n}(t)|^{2}$ in real space;
(b) a contour plot of $|\psi_{n}(t)|^{2}$;
(c) a contour plot of the phonon displacement $|\lambda_{n}(t)|$ in real space.} \label{D2-3}
\end{figure}

At variance with the single-parameter Merrifield Ansatz, the localized ${\rm D}_2$ trial state
employs two sets of variational parameters for independent descriptions of the two types of
constituting particles in the coupled exciton-phonon system.
By studying the outputs of time-dependent variational parameters, namely, the exciton
amplitude $\psi_{n} (t)$ and phonon displacement $\lambda_{n}(t)$, one is able to
probe the polaron dynamics which follows photo-excitation at a single site
of a one-dimensional molecular chain. At $t=0$, it is assumed that
the exciton is generated on one site and there are no initial
phonon displacements on the chain. Thanks to the exciton transfer integral and
the exciton-phonon coupling in the Holstein Hamiltonian, one witnesses the spreading of
the exciton amplitude and the growth of phonon deformations around the initial exciton site.
Here we use three examples to illustrate the time evolution of the exciton amplitude
and the phonon displacement of the ${\rm D}_2$ Ansatz in the site space.
As shown in Figs.~\ref{D2-1}, \ref{D2-2} and \ref{D2-3} for three sets of $(J,W)$,
$(0.5,0.8),(0.5,0.1)$ and $(1.0,0.8)$ respectively, the exciton amplitude propagates from the
initial site $n=0$ to the entire chain with a speed that is proportional to the transfer integral $J$.
The Huang-Rhys factor $S$ is set at $0.5$.
In Figs.~\ref{D2-1} and \ref{D2-2} ($J=0.5$), the first traveling wave of the exciton amplitude reaches the
the opposite end of the ring ($n=16$) at $t\approx2.6(2\pi/\omega_0)$, while in Fig.~\ref{D2-3} ($J=1.0$),
the exciton amplitude reaches the $n=16$ site at $t\approx1.3(2\pi/\omega_0)$.
This is because the propagation speed of the exciton, or equivalently, ${\partial E_{\rm ex}(k)}/{\partial k}$
with $E_{\rm ex}(k)$ the bare exciton band, is proportional to the exciton transfer integral $J$
according to Eq.~(\ref{H_ex}). The effect of phonon dispersion on the exciton is reflected in
the differing degree of phonon-induced dissipation in Figs.~\ref{D2-1}b and \ref{D2-2}b.
Exciton wave packets in Fig.~\ref{D2-2}b survive much longer than those in Fig.~\ref{D2-1}b
thanks to a much smaller phonon band width.

At $t=0$ there are no phonon deformations anywhere on the one-dimensional ring.
Due to photo-excitation at $t=0$, a high concentration of the exciton density at a single location,
appears in the vicinity of $n=0$ for a short time duration near $t=0$ [the red-colored spot near the point $(0,0)$ in
Figs.~\ref{D2-1}b, ~\ref{D2-2}b and ~\ref{D2-3}b], triggering
a pair of localized phonon wave packets which travel
at group velocities $\pm v_{q}$ with
\begin{equation}
v_{q}=\nabla_{q}\omega_{q}=\frac{2W}{\pi}.
\end{equation}
As shown in Figs.~\ref{D2-1}c, ~\ref{D2-2}c and ~\ref{D2-3}c, the angle between the two trajectories of
phonon wave packets departing from point $(0,0)$ but with opposite velocities in
Figs.~\ref{D2-1}c and ~\ref{D2-3}c ($W=0.8$) is about 8 times of that in Fig.~\ref{D2-2}c ($W=0.1$).

For a larger value of transfer integral, e.g., $J=1.0$ in Fig.~\ref{D2-3}, the left-moving and right-moving
wave packets of the exciton depart the site of creation and makes a quick rendezvous
at the opposite site of the ring ($n=16$), where the recombined exciton density
remains sufficiently high to trigger another pair of localized phonon wave packets
as is clearly demonstrated in Figs.~\ref{D2-3}b and \ref{D2-3}c. The exciton will
travel at a much reduced speed after the rendezvous at $n=16$, and further
stimulations of phonon wave packets are all but impossible for $J=1.0$.

\subsection{Vibrational Amplitude of Exciton and Phonon using $\tilde{\rm D}$ Ansatz}

\begin{figure}[t]
\begin{center}
\includegraphics[scale=0.6]{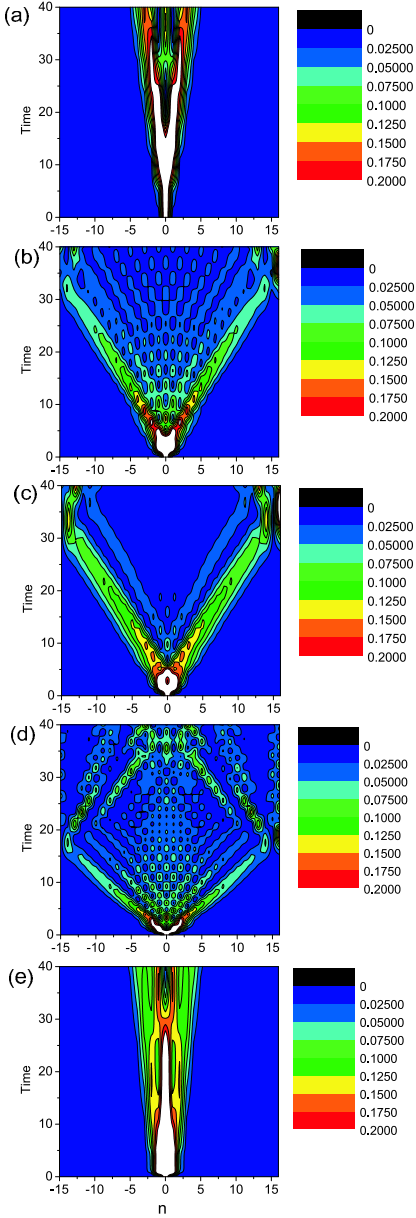}
\end{center}
\caption{Time evolution of the exciton probability $|\psi_{n}(t)|^{2}$ in real
space using $\tilde{\rm D}$ Ansatz. The unit of time is $1/\omega_0$.
(a) $J=0.1,~W=0.1,~S=0.5$; (b) $J=0.5,~W=0.8,~S=0.5$;
(c) $J=0.5,~W=0.1,~S=0.5$; (d) $J=1.0,~W=0.8,~S=0.5$;
(e) $J=1.0,~W=0.8,~S=2.0$.}\label{GL-1}
\end{figure}

\begin{figure}[t]
\begin{center}
\includegraphics[scale=0.583]{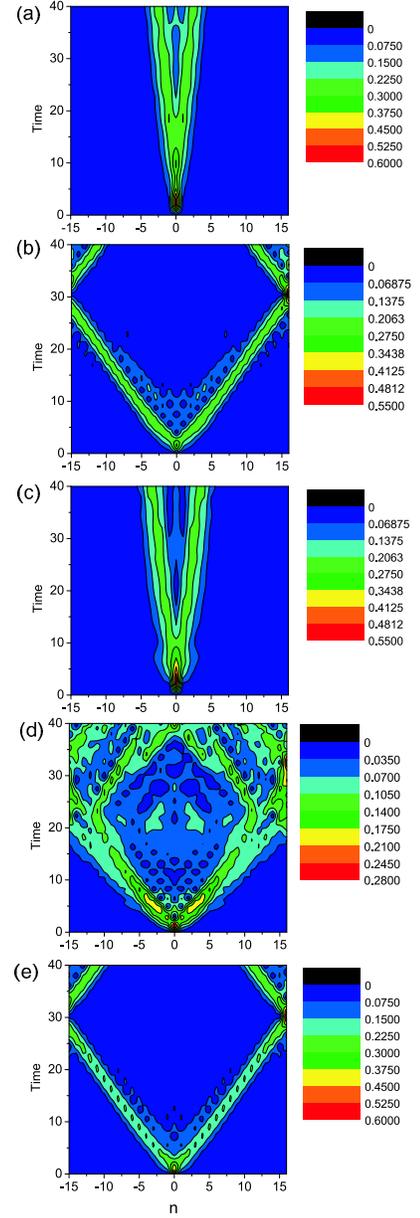}
\end{center}
\caption{Time evolution of the phonon displacements $|\lambda_{n}(t)|$
of the $\tilde{\rm D}$ Ansatz in real space. The unit of time is $1/\omega_0$.
(a) $J=0.1,~W=0.1,~S=0.5$; (b) $J=0.5,~W=0.8,~S=0.5$;
(c) $J=0.5,~W=0.1,~S=0.5$; (d) $J=1.0,~W=0.8,~S=0.5$;
(e) $J=1.0,~W=0.8,~S=2.0$.}\label{GL-2}
\end{figure}

\begin{figure}[tb]
\begin{center}
\includegraphics[scale=0.583]{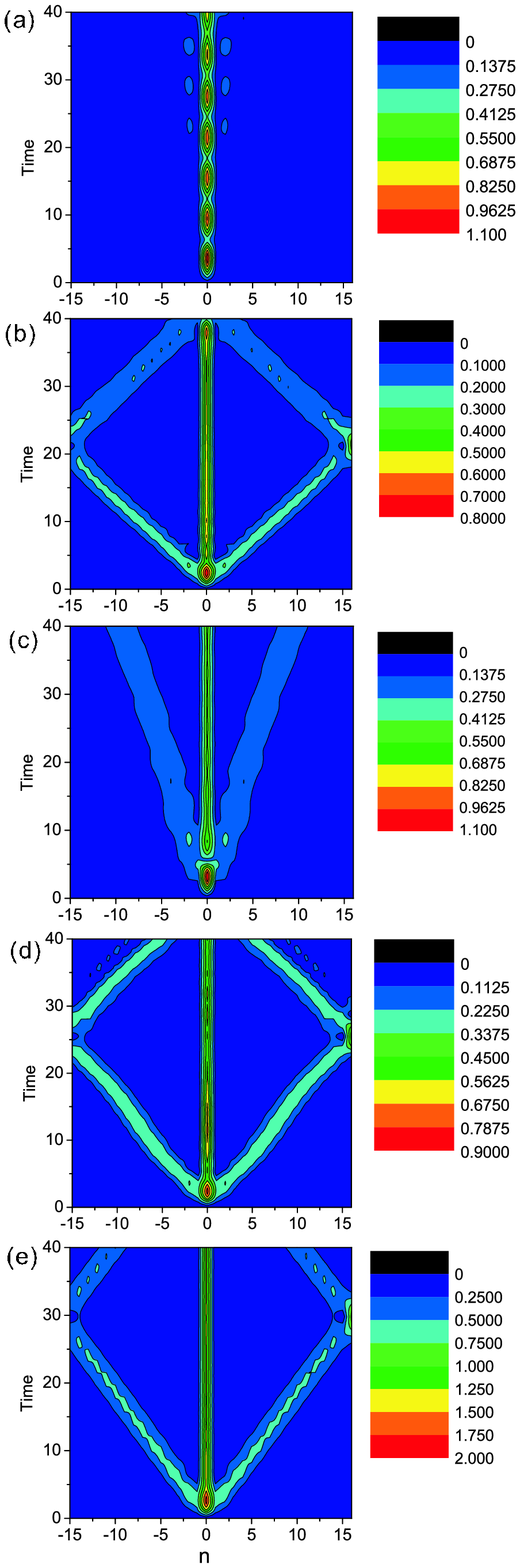}
\end{center}
\caption{Time evolution of the phonon vibrational amplitude $|\beta_{n}(t)|$ in real space
using $\tilde{\rm D}$ Ansatz. The unit of time is $1/\omega_0$.
(a) $J=0.1,~W=0.1,~S=0.5$; (b) $J=0.5,~W=0.8,~S=0.5$;
(c) $J=0.5,~W=0.1,~S=0.5$; (d) $J=1.0,~W=0.8,~S=0.5$;
(e) $J=1.0,~W=0.8,~S=2.0$.}\label{GL-3}
\end{figure}

In Figs.~\ref{GL-1}, \ref{GL-2} and \ref{GL-3}, time evolution of the variational parameters
$|\psi_{n}(t)|^{2}$, $|\lambda_{n}(t)|$ and $|\lambda_{n}(t)|$ is displayed for the $\tilde{\rm D}$ Ansatz.
When $J$ and $W$ are both small, the phonons are mostly localized in the vicinity of the site where
the exciton is initially created. The first set of phonon displacements, $|\lambda_{n}(t)|$,
will extend slightly to the neighboring sites as the time increases
because of a nonzero $J$ and a finite phonon dispersion (Fig.~\ref{GL-2}a).
But as shown in Fig.~\ref{GL-3}a, there is almost no spreading of $\beta_{n}(t)$,
suggesting that this portion of phonon displacements closely follows the exciton and
appears almost entirely on the site of exciton creation.
Fig.~\ref{GL-3}a also reveals oscillatory behavior of $|\beta_{n}(t)|$:
when $t$ equals $(2n+1)\pi/\omega_0$, $|\beta_{n}(t)|$ reaches its maximum value;
and when $t$ equals $2n\pi/\omega_0$, the amplitude is at its minimum.
Because the transfer integral $J$ is nearly zero, the exciton stays very close to
the site of creation. From Fig.~\ref{GL-1}a, it is obvious that the exciton
stays localized on one site with hardly any propagation for a small value of
transfer integral $J=0.1$. In this case, it seems that
oscillations of phonon amplitudes also influence the movement of the exciton.
The propagation of the exciton from the $n=0$ site to its nearest neighbors
occurs only at $t=2n\pi/\omega_0$ when the phonon amplitude reaches a minimum value.
Otherwise, because of the coupling with the phonons, the exciton
will be trapped and unable to transfer to other sites.

As $J$ is increased, the exciton will have the ability to move away from the site of creation.
As shown in Figs.~\ref{GL-1}b and \ref{GL-1}d, the larger $J$ is, the faster the exciton transfers.
Although $J$ has no direct influence over the phonon displacements,
a large $J$ induces more excitonic coherence between adjacent sites,
and in turn, causes more phonon deformations on those sites (cf. Fig.~\ref{GL-2}).
The propagation of the localized phonon wave packets related to $\beta_n(t)$
becomes faster $J$ is increased from $0.5$ to $1.0$ as shown in Fig.~\ref{GL-3}.

As for the $\tilde{\rm D}$ Ansatz, an increase in the width of the phonon dispersion $W$
will substantially deepen the phonon dissipative effect on
the exciton amplitude $\psi_{n}(t)$, and as a result the exciton coherence quickly
vanishes from the case of $W=0.8$. (cf. Fig.~\ref{GL-1}b and \ref{GL-1}c).
$W$ is also one main factor which determines the velocity of
the phonon wave packets for both types of phonon displacements $\lambda_n(t)$ and $\beta_n(t)$.
Fig.~\ref{GL-2}c shows that when $W$ is small, most of $\lambda_n(t)$ will be
localized on the $n=0$ site. As the phonon band width increases, $\lambda_n(t)$
will form two traveling wave packets with opposite directions and propagate to the entire chain as shown in
Fig.~\ref{GL-2}b.

\begin{figure}[tb]
\begin{center}
\includegraphics[scale=0.4]{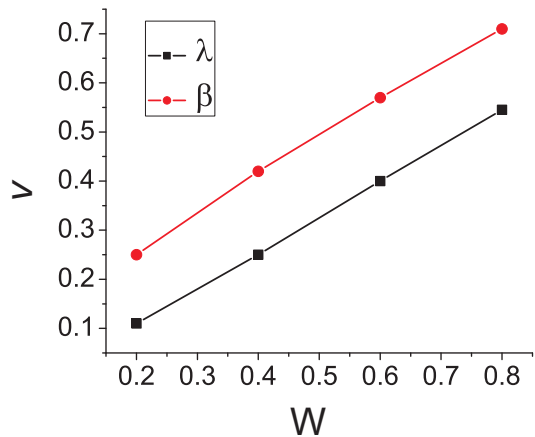}
\end{center}
\caption{Speeds of localized phonon wave packets as a function of the half width of
the phonon dispersion $W$ for the two sets of phonon displacements $\lambda_n(t)$ (in black)
and $\beta_n(t)$ (in red). $J=0.5,~S=0.5$.}\label{speed}
\end{figure}

The dynamics of $\beta_{n}(t)$ is more complex. As shown in Figs.~\ref{GL-2}b and \ref{GL-2}c,
the phonon displacements described by parameter $\lambda_{n}(t)$
propagate from the $n=0$ site to the rest of the aggregate.
But the same is not true for $\beta_{n}(t)$.
Most of $\beta_{n}(t)$ will stay localized on the site of exciton creation, but we can also find a small portion of
$\beta_{n}(t)$ propagating to the entire molecular chain as we increase
$W$ (cf. Fig.~\ref{GL-3}b and \ref{GL-3}c). From the definition of the $\tilde{\rm D}$ Ansatz,
Eq.~(\ref{DtildeDef}), $\beta_{n}(t)$ represents the portion of phonon displacements that
ride along the exciton, while $\lambda_{n}(t)$ labels the phonon displacements that are
not explicitly linked to the exciton. We therefore expect that the propagation speeds of
the localized phonon wave packets
from the two types of phonon displacements $\lambda_{n}(t)$ and $\beta_{n}(t)$ are different. But both of them
depend on the $W$ linearly, and the slopes are found to be the same, as shown in Fig.~\ref{speed}.

Lastly, we examine the effect of the Huang-Rhys factor $S$ which is a
dimensionless parameter representing the exciton-phonon coupling strength.
$S$ can be obtained directly from absorption and fluorescence spectra,
as it controls the vibrational progression that accompanies an exciton transition.
Comparing Fig.~\ref{GL-1}d with Fig.~\ref{GL-1}e, the exciton is found to be much
less mobile as $S$ is increased from $0.5$ to $2.0$, undergoing
the so-called self-trapping transition. The fact that Fig.~\ref{GL-1}d displays a much faster-moving polaron 
than Fig.~\ref{GL-1}e is also reflected in Fig.~\ref{GL-3}d, in which the localized phonon wave packet 
in $\beta_n (t)$ is found to be much longer-lasting than that in Fig.~\ref{GL-3}e, and its speed greater. 
In addition, Fig.~\ref{GL-2}d shows a few scattered localized phonon wave packets of $\lambda_n (t)$, while Fig.~\ref{GL-2}e
contains only a single localized phonon wave packet of $\lambda_n (t)$.

\subsection{Absorption Spectra and System Energies}

\begin{figure}[tb]
\begin{center}
\includegraphics[scale=0.8]{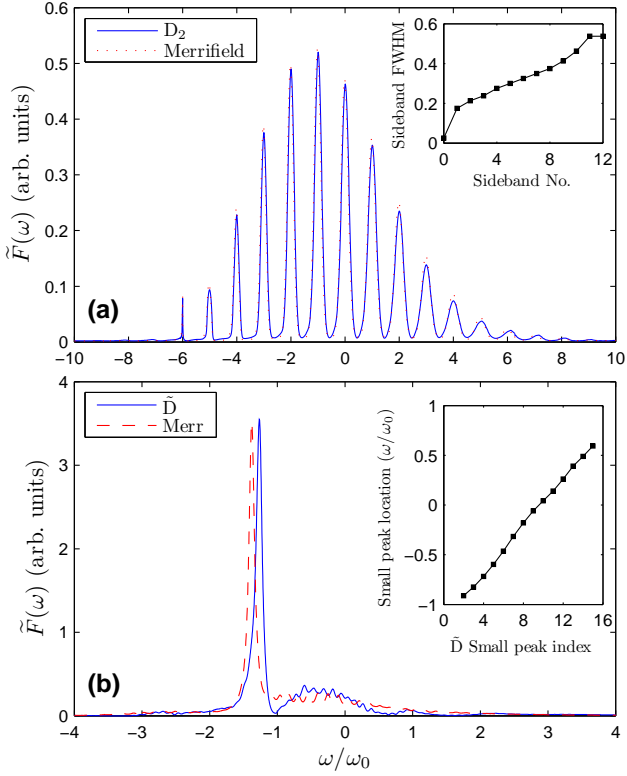}
\end{center}
\caption{Linear absorption spectra of a 32-site one-dimensional ring of a coupled exciton-phonon system.
(a) Using the ${\rm D}_2$ and Merrifield Ans\"{a}tze for $J=0.1,~W=0.1$ and $S=6.0$.
The decay factor used in the Fourier transformation is $0.01$.
(b) Using the $\tilde{\rm D}$ and Merrifield Ans\"{a}tze for $J=0.4,~W=0.8$ and $S=1.0$.
The decay factor used in the Fourier transformation is $0.05$.}\label{spectrum}
\end{figure}

Details on how to calculate linear absorption spectra of a one-dimensional
exciton-phonon system from the ${\rm D}_2$ and $\tilde{\rm D}$ Ans\"{a}tze can be found
in Appendix \ref{a4}. Fig.~\ref{spectrum}a and \ref{spectrum}b show two examples of the spectra get by
${\rm D}_2$ Ansatz and $\tilde{\rm D}$ Ansatz, respectively, and the results are compared with
those from the Merrifield Ansatz.
As shown in Fig.~\ref{spectrum}a, when $J$ is small (e.g., $J=0.1$) and $S$ is large (e.g., $S=6.0$),
the spectra obtained by the ${\rm D}_2$ and Merrifield Ans\"{a}tze are almost the same.
The phonon sidebands in Fig.~\ref{spectrum}a is labeled by $n=0,1,2,...$ from left to right.
The left most sideband, $n=0$, corresponding to the zero-phonon line,
is located at $\omega=-S\omega_0$ [Eq.~(\ref{relaxationE})]. For $S=6.0$,
the zero-phonon line is located at $\omega=-6\omega_0$ as shown in Fig.~\ref{spectrum}a.
According to the Huang-Rhys theory \cite{Huang}, the phonon sidebands at zero temperature
follow a Poisson distribution:
\begin{equation}
\widetilde{F}_{\rm abs}(\omega) = \exp(-S)\sum_{n=0}^\infty\frac{S^n}{n!}\delta(\omega+E_r-n\omega_0) \label{sideband}
\end{equation}
From Eq.~(\ref{sideband}), the tallest of phonon sidebands should be the
$n=S-1$ peak when $S\gg1$ and the overall distribution approaches a Gaussian.
Thus, for $S=6.0$, the tallest peak is at $n=5$ as shown in Fig.~\ref{spectrum}a.

Fig.~\ref{spectrum}b displays the absorption spectrum calculated from the $\tilde{\rm D}$
and Merrifield Ans\"{a}tze for $J=0.4,~W=0.8$ and $S=1.0$.
There are fine features with tiny peaks on the one-phonon manifold on the right side of
the zero-phonon line. These small peaks are attributed to the 32 phonon modes with
momenta $q=2\pi n_q/N~(N=32;~n_q=-15,-14,...,14,15,16)$.
As in this paper we assume a linear phonon dispersion as defined in Eq.~(\ref{width}),
there are altogether 17 values of $\omega_q$ for $N=32$.
Each value of $\omega_q$ should correspond to a small peak on the spectrum
if the coupling $g_q$ [Eq.~(\ref{interaction})] is not too small,
and from Eq.~(\ref{width}), one can obtain the locations of the small peaks
$\omega_{\rm sp}(n_q)~(n_q = 0,1,...,N/2)$:
\begin{equation}
\omega_{\rm sp}(n_q) = \omega_{\rm ZPL} + \omega_0 - W + \frac{4W}{N}n_q
\end{equation}
where $\omega_{\rm ZPL}$ is the location of the zero-phonon line. For $W=0.8\omega_0$ and $N=32$,
$\omega_{\rm sp}(n_q) = \omega_{\rm ZPL} + (0.2 + 0.1*n_q)\omega_0$ as shown
in the inset of Fig.~\ref{spectrum}b.

\begin{figure}[tb]
\begin{center}
\includegraphics[scale=0.78]{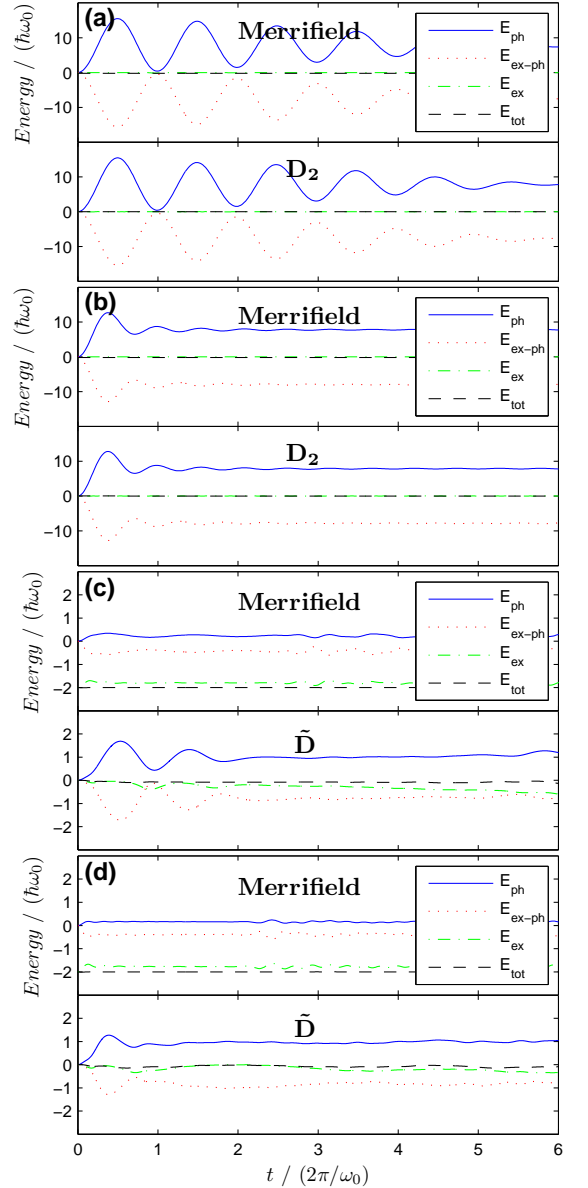}
\end{center}
\caption{System energies calculated from the ${\rm D}_2$ and $\tilde{\rm D}$ Ans\"{a}tz.
The results are compared with those of the Merrifield Ansatz.
(a) $J=0.1,~W=0.1,~S=4.0$; (b) $J=0.1,~W=0.8,~S=4.0$;
(c) $J=1.0,~W=0.1,~S=0.5$; (d) $J=1.0,~W=0.8,~S=0.5$.}\label{energy}
\end{figure}

Lastly, four cases of system energies calculated from the ${\rm D}_2$ and $\tilde{\rm D}$ Ans\"{a}tze
are shown in Fig.~\ref{energy} and compared with those from the Merrifield Ansatz .
Since the Hamiltonian defined in Eq.~(\ref{Hamiltonian}) is time-independent,
the total energy of the system is expected to be a constant during the time evolution.
At $t=0$, an exciton is assumed localized at one site and there are no phonons
on the entire ring. As shown in Fig.~\ref{energy}, the phonon energy $E_{\rm ph}$
rises from zero and oscillates as time goes on. In the meantime,
the interaction energy between the exciton and phonons, $E_{\rm ex-ph}$,
oscillates with almost the same amplitude but an opposite sign. And the total
energy $E_{\rm tot}$ of the system stays as a constant at all times. As shown in
Figs.~\ref{energy}a and \ref{energy}b, the ${\rm D}_2$ Ansatz and Merrifield Ansatz
agree with each other when $J$ is small and $S$ is large.
However, when $J$ is large (e.g., $J=1.0$) and $S$ is small (e.g., $S=0.5$),
these two Ans\"{a}tze no longer agree. For this case, the Merrifield Ansatz,
which is translationally invariant (a Bloch wave function) but with a built-in
small-polaron correlation, is not suitable to describe the dynamics.
As shown in Figs.~\ref{energy}c and \ref{energy}d,
the total system energy $E_{\rm tot}$ vanishes during the time evolution
of the $\tilde{\rm D}$ Ansatz. While in the Merrifield Ansatz, $E_{\rm tot}$ remains at
a constant value $-2$ that is equal to the initial value of $E_{\rm ex}$.
This is because in the Merrifield Ansatz, the exciton amplitude is assumed to
distribute uniformly over all the sites of the system, thus according to
Eq.~(\ref{H_ex}), $E_{\rm ex}$ has a negative value when $J$ is not negligible.
However, for an initial state in which the exciton is localized at one site,
according to Eq.~(\ref{H_ex}), $E_{\rm ex}$ should be zero at $t=0$ as
shown in Figs.~\ref{energy}c and \ref{energy}d for the $\tilde{\rm D}$ Ansatz.
This shows that the $\tilde{\rm D}$ Ansatz is a more flexible trial state
than the Merrifield Ansatz.

Lastly, we give a brief discussion on the validity of the Davydov Ans\"{a}tze.
A Davydov Ansatz is an approximative solution to the Schr\"{o}dinger equation with the Holstein
Hamiltonian. For a trial wave function $|D(t)\rangle$ that does not strictly obey
the Schr\"{o}dinger equation, the deviation vector $|\delta(t)\rangle$ can be written as
\begin{equation}\label{deltadef}
|\delta(t)\rangle \equiv i\hbar\frac{\partial}{\partial t}|D(t)\rangle - \hat{H}|D(t)\rangle
\end{equation}
Here $\hat{H}$ is the Holstein Hamiltonian  Eq.~(1).
For the Davydov ${\rm D}_1$ Ansatz, the explicit form of $|\delta(t)\rangle$ \cite{Davydov1}
was given by $\check{\rm S}$krinjar {\it et al.}~in 1988.
It was also proven that $|\delta(t)\rangle$ is orthogonal to $|D_1(t)\rangle$.
However, such orthogonality relations are insufficient to conclude that the deviation vector
$|\delta(t)\rangle$ is negligible, and the trial state is a good approximation to
the true solution of Schr\"{o}dinger equation. To have a quantitative measure of
the Schr\"{o}dinger-equation deviation,  one needs to calculate
the amplitude of the deviation vector $|\delta(t)\rangle$, which is defined as $\Delta(t)$:
\begin{equation}
\label{deltadef2}
\Delta(t) \equiv \sqrt{\langle\delta(t)|\delta(t)\rangle}
\end{equation}
An explicit expression for $\Delta(t)$ as the ${\rm D}_2$ Ansatz is substituted
into the Schr\"{o}dinger equation
is derived in Appendix \ref{a5}.

\begin{figure}[tb]
\begin{center}
\includegraphics[scale=0.68]{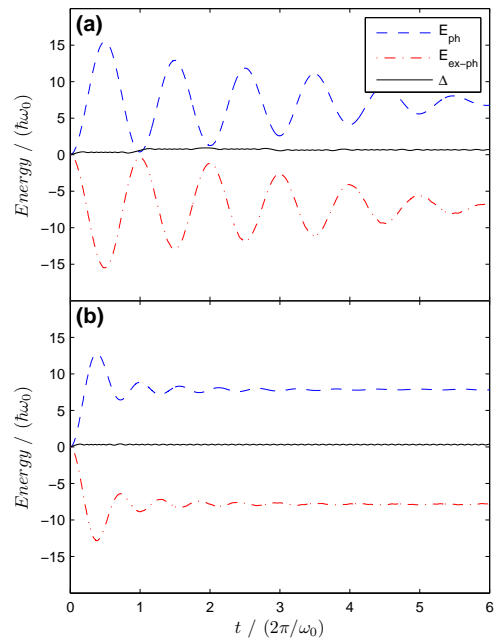}
\end{center}
\caption{Amplitude of the deviation vector $|\delta(t)\rangle$ calculated from the ${\rm D}_2$ Ansatz.
The results are compared with the system energies.
(a) $J=0.1,~W=0.1,~S=4.0$; (b) $J=0.1,~W=0.8,~S=4.0$.}\label{fig_deltaD2}
\end{figure}

Note that the dimension of $\Delta(t)$ is that of the energy. Therefore,
one can gauge whether the trial state is a good approximative solution by comparing $\Delta(t)$
with the system energies. Two examples of such comparisons for the ${\rm D}_2$ Ansatz are shown in
Figs.~\ref{fig_deltaD2}a and \ref{fig_deltaD2}b. The control parameters in Figs.~\ref{fig_deltaD2}a
and \ref{fig_deltaD2}b
are the same as those in Figs.~\ref{energy}a and \ref{energy}b, respectively.
For both two cases, the main system energies are $E_{\rm ph}$ and $E_{\rm ex-ph}$,
and $\Delta(t)$ is found to be negligible to either $E_{\rm ph}$ or $E_{\rm ex-ph}$.
We conclude that the ${\rm D}_2$ Ansatz yields quantitatively accurate solutions to the
Schr\"{o}dinger equation for these two cases.

\section{Conclusion}

In this paper we simulate polaronic dynamics in a one-dimensional
molecular chain following the Dirac-Frenkel time-dependent
variational approach. After the
optical excitation, the coupled exciton-phonon system will undergo a
relaxation process from the initial photo-induced nonequilibrium.
Based on the Holstein Hamiltonian, we examined time evolution of the exciton amplitude,
the reduced exciton density matrix,
the exciton coherence size, linear optical absorption, and induced phonon displacements for two types of
variational wave functions, namely, the Davydov ${\rm D}_2$  Ansatz, and a simplified variant of Davydov ${\rm D}_1$ Ansatz, also known as the $\tilde{\rm D}$ Ansatz. It is shown that following the equations of motion derived from the
time-dependent variation, the exciton amplitude will transfer from the site of creation
to neighboring sites, and as $J$ increases, the velocity of
exciton propagation will increase as well. Exciton-induced phonon deformations can be
found to form at the locations where the exciton has considerable densities, and then propagate to the entire aggregate
if there is sufficient phonon dispersion enabling mobility.
The half-width of the phonon dispersion, $W$,
determines the speeds of phonon wave packets because of a linear phonon dispersion relation given in the model.
For a large value of the transfer integral, the left-moving and right-moving
wave packets of the exciton departing from the site of creation may make a quick rendezvous
at the opposite end of the ring, and therefore, trigger a second pair of localized phonon wave packets.
Linear absorption spectra can be derived from the exciton amplitude and the phonon displacements computed from  the ${\rm D}_2$ and $\tilde{\rm D}$ Ans\"{a}tze. Various system energies are also calculated and analyzed for the one-dimensional
coupled exciton-phonon molecular ring after photo-excitation. Overall it is found that
$\tilde{\rm D}$ is a more flexible trial state, while the ${\rm D}_2$ Ansatz is rather efficient for computation, and its extension to higher spatial dimensions is a feasible generalization of our approach.

Our trial states in this paper are both localized wave functions, and we intend to work out detailed dynamics of their translationally invariant counterparts, the Toyozawa Ansatz and the GL Ansatz.  Our approaches here can be also readily extended to include other forms of exciton-phonon interactions, such as
asymmetric and symmetric off-diagonal coupling \cite{ZhaoOffDiagonal}, and higher-order couplings.
Work in this direction is now in progress.


\section*{Acknowledgments}
Support from the Singapore Ministry of Education through the
Academic Research Fund (Tier 2) under Project No. T207B1214 is
gratefully acknowledged.

\appendix

\section{The ${\rm D}_2$ trial state}
\label{a0}

The time evolution equations for ${\rm D}_1$, $\tilde{\rm D}$ and ${\rm D}_2$ Ans\"{a}tze
can be derived by employing Dirac-Frenkel time-dependent variation method.
We start from the simplest of the three, the ${\rm D}_2$ Ansatz. For
${\rm D}_2$ Ansatz, the Lagrangian is defined as
\begin{eqnarray}
L&=&\langle
D_{2}(t)|{\frac{i\hbar}{2}}\frac{\overset{\longleftrightarrow}{\partial}}{\partial
t}-\hat{H}|D_{2}(t)\rangle  \nonumber \\
&=& \frac{i\hbar}{2}[\langle
D_{2}(t)|\frac{\overrightarrow{\partial}}{\partial
t}|D_{2}(t)\rangle-\langle
D_{2}(t)|\frac{\overleftarrow{\partial}}{\partial
t}|D_{2}(t)\rangle] \nonumber \\
&&-\langle D_{2}(t)|\hat{H}|D_{2}(t)\rangle
\end{eqnarray}
where the first two terms connected with the time derivatives can be calculated as follows
\begin{eqnarray}
\langle D_{2}(t)|\frac{\overrightarrow{\partial}}{\partial
t}|D_{2}(t)\rangle
&=&\sum_{n}\psi_{n}^{*}\dot{\psi}_{n}-\sum_{n}|\psi_{n}|^{2}\sum_{q}\lambda_{q}^{*}\dot{\lambda}_{q}
\nonumber     \\
&&+\sum_{n}|\psi_{n}|^{2}\sum_{q}[-\frac{1}{2}(\dot{\lambda}_{q}\lambda_{q}^{*}+\lambda_{q}\dot{\lambda}_{q}^{*})], \nonumber    \\
\end{eqnarray}
and
\begin{eqnarray}
\langle D_{2}(t)|\frac{\overleftarrow{\partial}}{\partial
t}|D_{2}(t)\rangle&=&\sum_{n}\dot{\psi_{n}^{*}}\psi_{n}-\sum_{n}|\psi_{n}|^{2}\sum_{q}\dot{\lambda_{q}^{*}}\lambda_{q}
\nonumber    \\
&&+\sum_{n}|\psi_{n}|^{2}\sum_{q}[-\frac{1}{2}(\dot{\lambda}_{q}\lambda_{q}^{*}+\lambda_{q}\dot{\lambda}_{q}^{*})]\nonumber    \\
\end{eqnarray}
The remaining term is the average energy in the ${\rm D}_2$ trial state:
\begin{eqnarray}
\langle D_{2}|\hat{H}|D_{2}\rangle&=&\langle D_{2}|\hat{H}_{\rm ex}|D_{2}\rangle + \langle D_{2}|\hat{H}_{\rm ph}|D_{2}\rangle \nonumber \\
&& + \langle D_{2}|\hat{H}_{\rm ex-ph}|D_{2}\rangle
\end{eqnarray}
with
\begin{eqnarray}
\langle D_{2}|\hat{H}_{\rm ex}|D_{2}\rangle &=&
-J\sum_{n}\psi_{n}^{*}\psi_{n+1}
-J\sum_{n}\psi_{n}^{*}\psi_{n-1}
\nonumber\end{eqnarray}
\begin{equation}
\langle D_{2}|\hat{H}_{\rm
ph}|D_{2}\rangle=\sum_{n}|\psi_{n}|^{2}\sum_{q}\omega_{q}|\lambda_{q}|^{2}
\nonumber\end{equation}
\begin{equation}
\langle D_{2}|\hat{H}_{\rm
ex-ph}|D_{2}\rangle=\sum_{n}|\psi_{n}|^{2}\sum_{q}g_{q}\omega_{q}(\lambda_{q}e^{iqn}+\lambda_{q}^{*}e^{-iqn})
\nonumber
\end{equation}

Equations of motion from the trial wave function $D_{2}(t)$ are readily obtained from Eq.~(\ref{Lagrangian})
\begin{eqnarray}
-i\dot{\psi}_{n}(t)&=&\frac{i}{2}\psi_{n}(t)\sum_{q}[\dot{\lambda}_{q}(t)
\lambda_{q}^{*}(t)-\dot{\lambda}_{q}^{*}(t)\lambda_{q}(t)] \nonumber \\
&& +J\psi_{n+1}(t)+J\psi_{n-1}(t)\nonumber   \\
&&-\psi_{n}(t)\sum_{q}\omega_{q}|\lambda_{q}(t)|^{2}\nonumber   \\
&&-\psi_{n}(t)\sum_{q}g_{q}\omega_{q}[\lambda_{q}(t)e^{iqn}+\lambda_{q}^{*}(t)e^{-iqn}]\nonumber   \\
\label{alpha}
\end{eqnarray}
\begin{eqnarray}
-i\dot{\lambda}_{q}(t) &=&
-\sum_{n}|\psi_{n}|^{2}e^{-iqn}g_{q}\omega_{q}-\omega_{q}\lambda_{q}\label{beta}
\end{eqnarray}
It can easily be shown the norm of the ${\rm D}_2$ trial state is conserved, i.e.,
\begin{equation}
\frac{d}{dt}[\sum_{n=1}^{N}|\psi_{n}(t)|^{2}]=0
\end{equation}
And in this paper, we set
\begin{equation}
\sum_{n=1}^{N}|\psi_{n}(t)|^{2}=1
\end{equation}
Solutions to Eq.~(\ref{alpha}) and (\ref{beta}) provide the
dynamic information of the exciton-phonon system.

\section{The $\tilde{\rm D}$ trial state}
\label{a1}

For the $\tilde{\rm D}$ trial state, the localized backbone of the
translationally-invariant GL Ansatz, we can also derive coupled
equation for the time-dependent variational parameters using the
Lagrangian formalism
\begin{eqnarray}
L&=&\langle
\tilde{D}(t)|{\frac{i\hbar}{2}}\frac{\overset{\longleftrightarrow}{\partial}}{\partial t}-\hat{H}|\tilde{D}(t)\rangle  \nonumber \\
&=& \frac{i\hbar}{2}[\langle
\tilde{D}(t)|\frac{\overrightarrow{\partial}}{\partial
t}|\tilde{D}(t)\rangle-\langle
\tilde{D}(t)|\frac{\overleftarrow{\partial}}{\partial
t}|\tilde{D}(t)\rangle] \nonumber \\
&&-\langle \tilde{D}(t)|\hat{H}|\tilde{D}(t)\rangle
\label{LagrangianOfTildeD}
\end{eqnarray}
wherein the individual terms can be calculated as follows
\begin{eqnarray}
&&\langle \tilde{D}(t)|\frac{\overrightarrow{\partial}}{\partial
t}|\tilde{D}(t)\rangle\nonumber \\
&=&\sum_{n}\psi_{n}^{*}\dot{\psi}_{n}+\sum_{n}|\psi_{n}|^{2}[N^{-1}\sum_{q}(\dot{\beta}_{q}e^{-iqn}
\nonumber  \\
&&-\dot{\lambda}_{q})(\beta_{q}^{*}e^{iqn}-\lambda_{q}^{*})]+\sum_{n}|\psi_{n}|^{2} \nonumber  \\
&&[-\frac{1}{2}N^{-1}\sum_{q}(\dot{\beta}_{q}e^{-iqn}
-\dot{\lambda}_{q})(\beta_{q}^{*}e^{iqn}-\lambda_{q}^{*})  \nonumber   \\
&&+(\beta_{q}e^{-iqn}-\lambda)(\dot{\beta}_{q}^{*}e^{iqn}-\dot{\lambda}_{q})]
\end{eqnarray}
\begin{eqnarray}
&&\langle \tilde{D}(t)|\frac{\overleftarrow{\partial}}{\partial
t}|\tilde{D}(t)\rangle\nonumber \\&=&\sum_{n}\dot{\psi}_{n}^{*}\psi_{n}
+\sum_{n}|\psi_{n}|^{2}[N^{-1}\sum_{q}(\dot{\beta}_{q}^{*}e^{iqn}    \nonumber  \\
&&-\dot{\lambda}_{q}^{*})
(\beta_{q}e^{-iqn}-\lambda_{q})]+\sum_{n}|\psi_{n}|^{2}   \nonumber \\
&&[-\frac{1}{2}N^{-1}\sum_{q}(\dot{\beta}_{q}e^{-iqn}-\dot{\lambda}_{q})(\beta_{q}^{*}e^{iqn}-\lambda_{q}^{*})
\nonumber  \\
&&+(\beta_{q}e^{-iqn}-\lambda)(\dot{\beta}_{q}^{*}e^{iqn}-\dot{\lambda}_{q})]
\end{eqnarray}
and
\begin{eqnarray}
\langle \tilde{D}|\hat{H}|\tilde{D}\rangle
&=&\langle \tilde{D}|\hat{H}_{\rm ex}|\tilde{D}\rangle+\langle \tilde{D}|\hat{H}_{\rm ph}|\tilde{D}\rangle
+\langle \tilde{D}|\hat{H}_{\rm ex-ph}|\tilde{D}\rangle  \nonumber  \\
\label{HamiltonianOfTildeD}
\end{eqnarray}
with
\begin{eqnarray}
\langle \tilde{D}|\hat{H}_{\rm ex}|\tilde{D}\rangle =
-J\sum_{n}\psi_{n}^{*}(\psi_{n+1}S_{n,n+1}+\psi_{n-1}S_{n,n-1})  \nonumber
\end{eqnarray}
\begin{eqnarray}
\langle \tilde{D}|\hat{H}_{\rm
ph}|\tilde{D}\rangle=N^{-1}\sum_{n}|\psi_{n}|^{2}\sum_{q}\omega_{q}|\beta_{q}e^{-iqn}-\lambda_{q}|^{2}
\nonumber
\end{eqnarray}
\begin{eqnarray}
\langle \tilde{D}|\hat{H}_{\rm
ex-ph}|\tilde{D}\rangle&=&N^{-1/2}\sum_{n}|\psi_{n}|^{2}\sum_{q}g_{q}\omega_{q}[(\beta_{q}^{*}e^{iqn}-\lambda_{q}^{*})
\nonumber \\
&&e^{-iqn}+(\beta_{q} e^{-iqn}-\lambda_{q}^{*})e^{-iqn}].   \nonumber
\end{eqnarray}

Substituting Eqs.~(\ref{LagrangianOfTildeD})-(\ref{HamiltonianOfTildeD}) into Eq.~(\ref{Lagrangian}),
one arrives at the equations of time evolution for the $\tilde{\rm D}$ Ansatz:
\begin{eqnarray}
-i\dot{\psi}_{n}(t)&=&\frac{i}{2}N^{-1}\psi_{n}\sum_{q}[(\dot{\beta_{q}}e^{-iqn}-\dot{\lambda_{q}})
(\beta_{q}^{*}e^{iqn}-\lambda_{q}^{*})  \nonumber   \\
&&-(\dot{\beta_{q}^{*}}e^{iqn}-\dot{\lambda_{q}^{*}})(\beta_{q}e^{-iqn}-\lambda_{q})]   \nonumber   \\
&& +J\psi_{n+1}S_{n,n+1}+J\psi_{n-1}S_{n,n-1}   \nonumber   \\
&&-N^{-1}\psi_{n}\sum_{q}\omega_{q}|\beta_{q}e^{-iqn}-\lambda_{q}|^{2}-N^{-1/2}\psi_{n}  \nonumber  \\
&&\sum_{q}g_{q}\omega_{q}[\beta_{q}^{*}-\lambda_{q}^{*}e^{iqn}+\beta_{q}-\lambda_{q}e^{iqn}]
\label{GL_1}
\end{eqnarray}
\begin{eqnarray}
-iN^{-1}&&\sum_{n}|\psi_{n}|^{2}\dot{\lambda}_{q}(t)=\nonumber \\
&& -i\sum_{n}|\psi_{n}|^{2}\dot{\beta}_{q}e^{-iqn}N^{-1} \nonumber  \\
&& +\frac{1}{2}N^{-1}J\sum_{n}\psi_{n}^{*}\psi_{n+1}S_{n,n+1}  \nonumber  \\
&&\beta_{q}[e^{-iqn}-e^{-iq(n+1)}]   \nonumber   \\
&& +\frac{1}{2}N^{-1}J\sum_{n}\psi_{n}^{*}\psi_{n-1}S_{n,n-1}   \nonumber  \\
&&\beta_{q}[e^{-iqn}-e^{-iq(n-1)}]   \nonumber   \\
&&+N^{-1}\sum_{n}|\psi_{n}|^{2}\omega_{q}(\beta_{q}e^{-iqn}-\lambda_{q})    \nonumber  \\
&&+N^{-1/2}\sum_{n}|\psi_{n}|^{2}g_{q}\omega_{q}e^{-iqn}
\label{GL_2}
\end{eqnarray}
\begin{eqnarray}
-iN^{-1}&& \sum_{n}|\psi_{n}|^{2}\dot{\beta}_{q}= \nonumber    \\
&&-iN^{-1}\sum_{n}|\psi_{n}|^{2}\dot{\lambda}_{q}e^{iqn}   \nonumber    \\
&& +JN^{-1}\sum_{n}\psi_{n}^{*}\psi_{n+1}S_{n,n+1}   \nonumber  \\
&&[\beta_{q}(e^{-iq}-1)+\frac{1}{2}\lambda_{q}e^{iqn}(e^{iq}-1)]
\nonumber  \\
&& +JN^{-1}\sum_{n}\psi_{n}^{*}\psi_{n-1}S_{n,n-1}  \nonumber   \\
&&[\beta_{q}(e^{iq}-1)+\frac{1}{2}\lambda_{q}e^{iqn}(e^{-iq}-1)]
\nonumber  \\
&&-N^{-1}\sum_{n}|\psi_{n}|^{2}\omega_{q}(\beta_{q}-\lambda_{q}e^{iqn})   \nonumber \\
&&-N^{-1/2}\sum_{n}|\psi_{n}|^{2}g_{q}\omega_{q}
\label{GL_3}
\end{eqnarray}

\section{The ${\rm D}_1$ trial state}
\label{a2}

The same procedure in Appendices \ref{a0} and \ref{a1} can be
applied to the ${\rm D}_1$ Ansatz \cite{ZhaoDavydov} with variational parameters ${\psi}_{n}$
and ${\lambda}_{qn}$, and one can obtain
\begin{eqnarray}
-i\dot{\psi}_{n}(t)&=&\frac{i}{2}N^{-1}\psi_{n}\sum_{q}[\dot{\lambda}_{qn}\lambda_{qn}^{*}-c.c.] \nonumber \\
&& +J\psi_{n+1}S_{n,n+1}+J\psi_{n-1}S_{n,n-1}   \nonumber   \\
&&-N^{-1/2}\psi_{n}\sum_{q}g_{q}\omega_{q}[\lambda_{qn}e^{iqn} + c.c.]  \nonumber  \\
&&-N^{-1}\psi_{n}\sum_{q}\omega_{q}|\lambda_{qn}|^{2}
\end{eqnarray}
\begin{eqnarray}
-iN^{-1}&&|\psi_{n}|^{2}\dot{\lambda}_{qn}(t)=\nonumber \\
&& + N^{-1}J\psi_{n}^{*}\psi_{n+1}(\lambda_{q,n+1}-\lambda_{qn})S_{n,n+1}  \nonumber  \\
&& + N^{-1}J\psi_{n}^{*}\psi_{n-1}(\lambda_{q,n-1}-\lambda_{qn})S_{n,n-1}   \nonumber  \\
&& - N^{-1}|\psi_{n}|^{2}\omega_{q}\lambda_{qn}    \nonumber  \\
&& - N^{-1/2}|\psi_{n}|^{2}g_{q}\omega_{q}e^{-iqn}
\end{eqnarray}

These coupled differential equations can be numerically solved by
the fourth-order Runge-Kutta method, and work on this is now in
progress.

\section{Numerical Details}
\label{a3}

There are various numerical approaches to solve coupled differential equations
such as Eqs.~(\ref{alpha}) and (\ref{beta}) for the ${\rm D}_2$ Ansatz and
Eqs.~(\ref{GL_1}), (\ref{GL_2}) and (\ref{GL_3}) for the $\tilde{\rm D}$ Ansatz.
One way is to transform these equations into the Volterra integral equations \cite{Volterra}
for minimization purposes, and then use a nonlinear optimization method such as
the Newton-Raphson method to solve them. Another approach is to solve the time-dependent
differential equations directly. The latter method has a much higher computational efficiency,
but it requires a higher precision in the single iterative time step since the computational error
may accumulate as the number of the iterative steps increases.

In this paper, we use the Runge-Kutta fourth-order method \cite{Runge1, Runge2} to solve
Eqs.~(\ref{alpha}) and (\ref{beta}) and Eqs.~(\ref{GL_1}), (\ref{GL_2}) and (\ref{GL_3}).
The Runge-Kutta fourth-order method is widely used to solve differential equations.
Its single step error is fifth order, i.e., $O(\Delta t^{5})$.
The algorithm for this method can be described as follows:
\begin{eqnarray}
k_{1}&=&f(t_{n},\alpha(t_{n}))\Delta t   \nonumber \\
k_{2}&=&f(t_{n}+\frac{1}{2}\Delta t,\alpha(t_{n})+\frac{1}{2}k_{1})\Delta t   \nonumber  \\
k_{3}&=&f(t_{n}+\frac{1}{2}\Delta t,\alpha(t_{n})+\frac{1}{2}k_{2})\Delta t   \nonumber  \\
k_{4}&=&f(t_{n}+\Delta t,\alpha(t_{n})+k_{3})\Delta t   \nonumber   \\
\alpha(t_{n+1})&=&\alpha(t_{n})+\frac{k_{1}}{6}+\frac{k_{2}}{3}+\frac{k_{3}}{3}+\frac{k_{4}}{6}+O(\Delta t^{5}) \nonumber
\end{eqnarray}
where
\begin{equation}
f(t_{n},\alpha(t_{n}))=\frac{d\alpha(t_{n})}{dt_{n}}\nonumber
\end{equation}
and from $t_{n}$ to $t_{n+1}\equiv
t_{n}+\Delta t $.

For the ${\rm D}_2$ Ansatz, $\alpha$ represents $\psi_{n}$ or $\lambda_{q}$, and is
obtained by Eqs.~(\ref{alpha}) and (\ref{beta}). Then, by calculating $k_{1}$, $k_{2}$,
$k_{3}$ and $k_{4}$, one can get the parameters $\alpha(t_{n+1})$ for the next time step
using the fourth-order Runge-Kutta method.

\section{Linear absorption spectrum by ${\rm D}_2$ and $\tilde{\rm D}$ Ans\"{a}tze}
\label{a4}

The linear absorption spectrum of the exciton-phonon system studied in
this paper is calculated by
\begin{equation}
\widetilde{F}(\omega)=\frac{1}{\pi}{\rm Re}\int_{0}^{\infty}F(t)e^{i\omega t}dt
\end{equation}
with
\begin{eqnarray}\label{Ft}
F(t) =~_{\rm ph}\langle0|_{\rm ex}\langle0|\hat{P}e^{-i\hat{H}t}\hat{P}^{\dagger}|0\rangle_{\rm ex}|0\rangle_{\rm ph}
\end{eqnarray}
where $\hat{P}$ is the polarization operator
\begin{equation}\label{polarization}
\hat{P} = \mu \sum_{n}(|n\rangle_{\rm ex}~_{\rm ex}\langle0| + |0\rangle_{\rm ex}~_{\rm ex}\langle n|)
\end{equation}
Here $\mu$ is the transition dipole matrix element for a single
site, and $|n\rangle_{\rm ex}$ is the exciton state at the $n$th
site $|n\rangle_{\rm ex} \equiv \hat{a}_{n}^{\dagger}|0\rangle_{\rm ex}$.

Substituting Eq.~(\ref{polarization}) into Eq.~(\ref{Ft}) one obtains
\begin{equation}\label{Ft2}
F(t) = \mu^2 \sum_n\sum_m~_{\rm ph}\langle0|_{\rm ex}\langle m|e^{-i\hat{H}t}|n\rangle_{\rm ex}|0\rangle_{\rm ph}
\end{equation}
wherein $e^{-i\hat{H}t}|n\rangle_{\rm ex}|0\rangle_{\rm ph}$
can be approximated by a Davydov trial state, for example, by a ${\rm D}_2$ trial state:
\begin{eqnarray}\label{D2Approx}
&&e^{-i\hat{H}t}|n\rangle_{\rm ex}|0\rangle_{\rm ph} \approx \nonumber\\
&&\sum_{n'}\psi_{n'-n}(t)|n'\rangle_{\rm ex}\nonumber\\
&&\times\exp(\sum_q[\lambda_q(t)\hat{b}_q^{\dagger}-\lambda_q^*(t)\hat{b}_q])|0\rangle_{\rm ph}
\end{eqnarray}
where the variational parameters $\psi_{n'-n}(t)~(n'=0,2,...,N-1)$ and $\lambda_q(t) (q=-N/2+1,-N/2+2,...,N/2)$
have the following initial values
\begin{equation}\label{psi0}
\psi_{n'-n}( t=0) = \delta_{n'-n},
\end{equation}
and
\begin{equation}\label{lambda0}
\lambda_q(t=0) = 0.
\end{equation}
Substituting Eq.~(\ref{D2Approx}) into Eq.~(\ref{Ft2}), we obtain the formula to calculate $F(t)$ by ${\rm D}_2$ Ansatz:
\begin{eqnarray}\label{FtD2}
F(t) &=& \mu^2\sum_n\sum_m\psi_m(t) \nonumber\\
&&\times~_{\rm ph}\langle0|\exp(\sum_q[\lambda_q(t)\hat{b}_q^{\dagger}-\lambda_q^*(t)\hat{b}_q])|0\rangle_{\rm ph} \nonumber\\
&=& \mu^2\sum_n\sum_m\psi_m(t)\exp(-\frac{1}{2}\sum_q|\lambda_q(t)|^2) \nonumber\\
&=& \mu^2 N\sum_m\psi_m(t)\exp(-\frac{1}{2}\sum_q|\lambda_q(t)|^2)
\end{eqnarray}
wherein $\lambda_q(t)$ and $\psi_m(t)$ are initialized by Eq.~(\ref{lambda0}) and
Eq.~(\ref{psi0}) (i.e., $\psi_m(0)=\delta_{m,0}$) and then solved by Eqs.~(\ref{beta}) and (\ref{alpha}).

The same procedure from Eq.(\ref{D2Approx}) to (\ref{FtD2}) can be
applied to $\tilde{\rm D}$ Ansatz, and one obtains
\begin{eqnarray}\label{FtDtilde}
F(t)=\mu^2 N\sum_m\psi_m(t)\exp(-\frac{1}{2}\sum_q|\beta_q(t)e^{-iqm}-\lambda_q(t)|^2)\nonumber\\
\end{eqnarray}
where the variational parameters $\psi_m(t)$, $\beta_q(t)$ and $\lambda_q(t)$
are solved by Eqs.~(\ref{GL_1})-(\ref{GL_3}) with the initial values
$\psi_m(0)=\delta_{m,0}$, $\beta_q(0)=0$ and $\lambda_q(0)=0$.

\section{Deviation vector of the ${\rm D}_2$ Ansatz}
\label{a5}

Substituting Eqs.~(\ref{Hamiltonian})-(\ref{ex-ph}), (\ref{D2def}) and (\ref{D2lambdadef})
into Eq.~(\ref{deltadef}), one obtains the expression of $|\delta(t)\rangle$ for the ${\rm D}_2$ Ansatz:
\begin{eqnarray}\label{deltaD2}
|\delta(t)\rangle &=& \sum_n\hat{B}_n^{\dagger}\{i\{\dot{\psi}_n(t)+ \nonumber\\
&&\psi_n(t)\sum_q\{\hat{b}_q^{\dagger}\dot{\lambda}_q(t) - {\rm Re}[\dot{\lambda}_q(t)\lambda_q^*(t)]\}\}\nonumber\\
&&+J[\psi_{n+1}(t)+\psi_{n-1}(t)] - \psi_n(t)\sum_q\omega_q\hat{b}_q^{\dagger}\lambda_q(t) \nonumber\\
&&+\psi_n(t)\sum_qg_q\omega_q[\hat{b}_q^{\dagger}e^{-inq} + \lambda_q(t)e^{inq}]\} \nonumber\\
&&\exp(\sum_q[\lambda_q(t)\hat{b}_q^{\dagger}-\lambda_q^*(t)\hat{b}_q])|0\rangle_{\rm ph}|0\rangle_{\rm ex}
\end{eqnarray}
Then, the expression of $\langle\delta(t)|\delta(t)\rangle$ for the ${\rm D}_2$ Ansatz can be obtained:
\begin{eqnarray}\label{delta2D2}
&&\langle\delta(t)|\delta(t)\rangle = J^2\sum_n|\psi_{n+1}(t) + \psi_{n-1}(t)|^2 \nonumber\\
&&-2J[\sum_q\omega_q|\lambda_q(t)|^2]{\rm Re}[\sum_n[\psi_{n+1}^*(t) + \psi_{n-1}^*(t)]\psi_n(t)] \nonumber\\
&&+4J{\rm Re}[\sum_n[\psi_{n+1}^*(t) + \psi_{n-1}^*(t)]\theta_n(t)] \nonumber\\
&&-2J{\rm Re}[\sum_n[\psi_{n+1}^*(t) + \psi_{n-1}^*(t)]\vartheta_n(t)] \nonumber\\
&&+[\sum_n|\psi_n(t)|^2]\{[\sum_q\omega_q|\lambda_q(t)|^2]^2 + \sum_q|\omega_q\lambda_q(t)|^2\} \nonumber\\
&&-4[\sum_q\omega_q|\lambda_q(t)|^2]^2\sum_n\psi_n^*(t)\theta_n(t) \nonumber\\
&&-2\sum_n\psi_n^*(t)\psi_n(t)\sum_qg_q\omega_q^2{\rm Re}[\lambda_q(t)e^{inq}] \nonumber\\
&&+2[\sum_q\omega_q|\lambda_q(t)|^2]{\rm Re}[\sum_n\psi_n^*(t)\vartheta_n(t)] \nonumber\\
&&+2[\sum_n|\psi_n(t)|^2]{\rm Im}[\sum_q\omega_q\dot{\lambda}_q(t)\lambda_q^*(t)] \nonumber\\
&&+4\sum_n|\theta_n(t)|^2 + [\sum_n|\psi_n(t)|^2]\sum_q(g_q\omega_q)^2 \nonumber\\
&&-4{\rm Re}[\sum_n\theta_n^*(t)\vartheta_n(t)] \nonumber\\
&&-2{\rm Im}[\sum_n\psi_n^*(t)\psi_n(t)\sum_qg_q\omega_q\dot{\lambda}_q(t)e^{inq}] \nonumber\\
&&+\sum_n|\vartheta_n(t)|^2 + [\sum_n|\psi_n(t)|^2]\sum_q|\dot{\lambda}_q(t)|^2
\end{eqnarray}
where
\begin{equation}
\theta_n(t) \equiv \psi_n(t)\sum_qg_q\omega_q{\rm Re}[\lambda_q(t)e^{inq}]
\end{equation}
and
\begin{equation}
\vartheta_n(t) \equiv \psi_n(t){\rm Im}[\sum_q\dot{\lambda}_q(t)\lambda_q^*(t)] - i\dot{\psi}_n(t)
\end{equation}

\end{document}